\documentclass[aps,prl,twocolumn,groupedaddress,showpacs,floatfix]{revtex4}

\usepackage{graphicx}
\usepackage{subfigure}
\usepackage{amsmath}
\def\ket#1{|#1\rangle}

\begin{document}
	
\author{N. Arunkumar$^{1}$, A. Jagannathan$^{1,2}$, and J. E. Thomas$^{1}$}
\affiliation{$^{1}$Department of  Physics, North Carolina State University, Raleigh, NC 27695}
\affiliation{$^{2}$Department of Physics, Duke University, Durham, NC 27708}
\date{\today}

\title{Designer Spatial Control of Interactions in Ultracold Gases}

\begin{abstract}

Designer optical control of interactions in ultracold atomic gases has wide application, from creating new quantum phases to modeling the physics of black holes. We demonstrate spatial control of interactions in a two-component cloud of $^6$Li fermions, using electromagnetically induced transparency (EIT) to create a ``sandwich" of resonantly and weakly interacting regions. Interaction designs are imprinted on the trapped cloud by two laser beams and manipulated with just MHz changes in the frequency of one beam. We employ radio-frequency spectroscopy to measure the imprinted 1D spatial profiles of the local mean-field interactions and to demonstrate that the tuning range of the scattering length is the same for both optical and magnetic control. All of the data are in excellent agreement with our continuum-dressed state theoretical model of optical control, which includes both the spatial and momentum dependence of the interactions.

\end{abstract}

\maketitle

Tunability of interactions in ultracold atomic gases has been achieved by exploiting magnetically controlled collisional (Feshbach) resonances~\cite{ChinFeshbach}, where the total energy of two colliding atoms in an energetically open channel is tuned into resonance with a bound dimer state in a closed channel. Optical field control offers a much richer palate, by creating designer interactions with high resolution in position, energy, momentum, and time. These techniques enable new paradigms. For example, energy resolution will provide  better models of neutron matter by controlling the effective range~\cite{WuOptControl1, WuOptControl2}, while momentum resolution will permit non-zero momentum pairing in two component Fermi gases,  i.e., synthetic FFLO states~\cite{OptcontrolCOM, FuldeFerrell}. The increased temporal resolution enables studies of non-equilibrium thermodynamics of strongly interacting gases on time scales faster than the Fermi time~\cite{BulgacYoon09}. Spatial manipulation of interactions can be utilized to study controllable soliton emission~\cite{SolitonEmission}, exotic quantum phases~\cite{QuantumPhases}, long-living Bloch oscillations of matter waves~\cite{BlochOscillations}, the physics of Hawking radiation from black holes~\cite{HawkingBlackhole}, and scale-invariant dimer pairing~\cite{DeanLeeScaleInv}. However, optical techniques generally suffer from atom loss and heating due to spontaneous scattering, which severely limits their applicability~\cite{WalravenOptTuning, JulienneOptTuning, PLettOFB, EnomotoOFB, TheisOFB, YamazakiSpatialMod, RempeOptControl,CetinaOptical, ChinMagicOptControl}.

In a major breakthrough for suppressing spontaneous scattering, Bauer et al.,~\cite{RempeOptControl} used a bound-to-bound transition in the closed channel, which is far away from the atomic resonance. To further suppress atom loss, large detunings on the bound-bound transition were employed. The large detunings limited the tunability of the scattering length $a$ to $\Delta a\simeq 2\,a_{bg}$, where $a_{bg}$ is the background scattering length. In addition, interactions were tuned by changing the intensity of the laser light, which changes the net external potential experienced by the atoms.
Recently, Clark and coworkers~\cite{ChinMagicOptControl} avoided this problem by using a ``magic" wavelength, tuned in between $D1$ and $D2$ lines of $^{137}$Cs atoms, to suppress the atomic polarizability and hence the change in the external potential, but achieved a tunability of only $\simeq 0.2\,a_{bg}$. Further, this technique cannot be adopted universally, as it leads to excessive atom loss in atomic species such as  $^6$Li, where the $D1$ and $D2$ lines are closely spaced.

\begin{figure*}[htb]
	\includegraphics[height = 2.2 in]{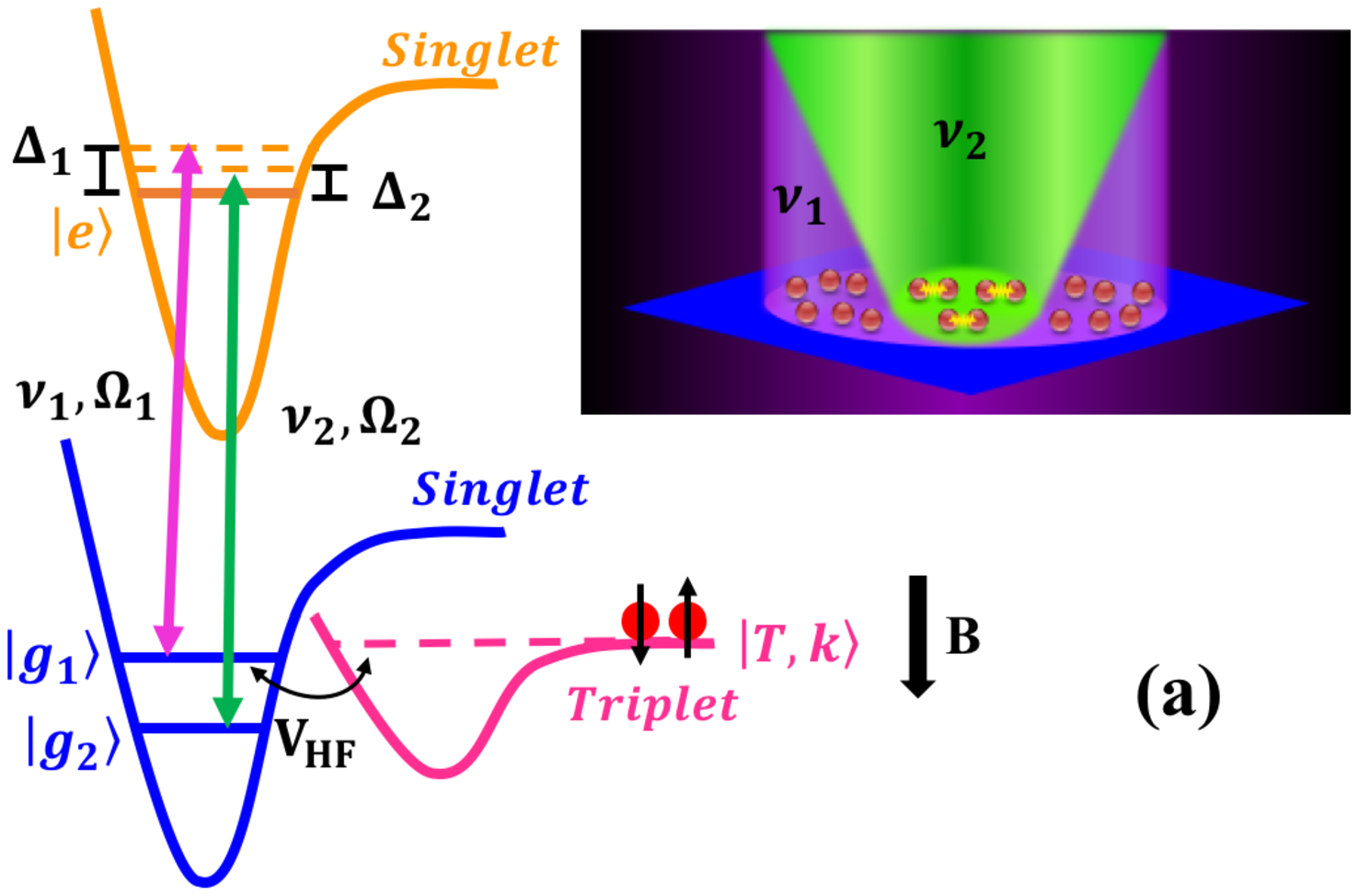}\hspace{0.125in}\includegraphics[height = 2.2 in]{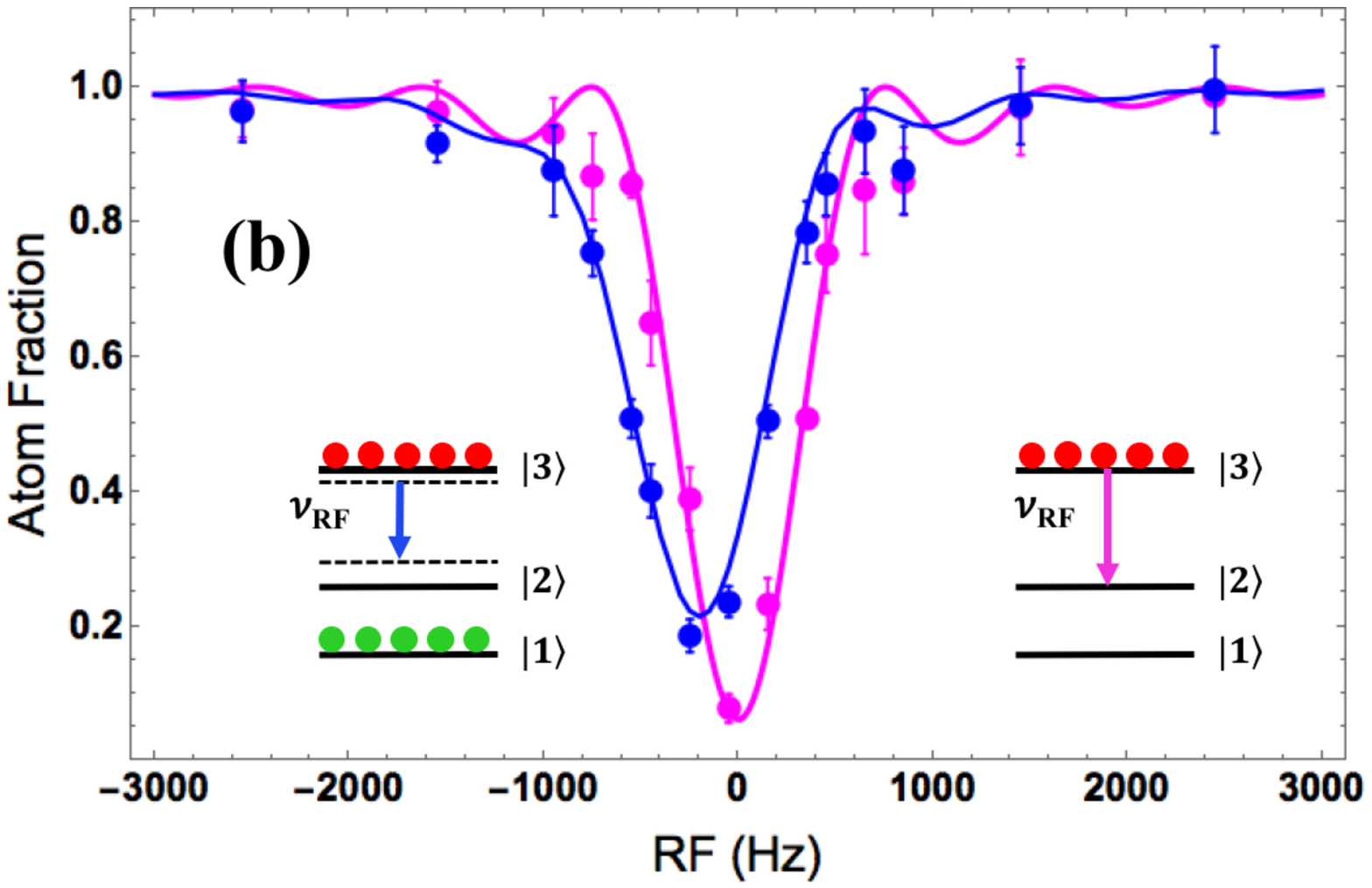}
	\caption{Basic level scheme to control interactions using electromagnetically induced transparency (EIT).	
		(a) Optical fields $\nu_1$ (Rabi frequency $\Omega_1$ and detuning $\Delta_1$)  and $\nu_2$ (Rabi frequency $\Omega_2$ and detuning $\Delta_2$)  couple the ground molecular states $\ket{g_1}$ and $\ket{g_2}$ to the excited molecular state $\ket{e}$ of the singlet potential, allowing precise tuning of the state $\ket{g_1}$ from below ($\delta <0$) to above ($\delta >0$) its unshifted position, where $\delta = \Delta_2 - \Delta_1 $ is the two-photon detuning. Inset shows the optical field arrangement for creating an interaction ``sandwich." The central region of the atomic cloud illuminated by both $\nu_1$ and $\nu_2$ beams are resonantly interacting. The outer regions of the atomic cloud illuminated $\textit{only}$ by the $\nu_1$ beam are weakly interacting. (b) Measuring mean-field interactions. RF spectra of atoms transferred from hyperfine state $\ket{3}$ to $\ket{2}$, obtained with (blue) and without (magenta) atoms present in state $\ket{1}$. Inset shows the mean-field induced energy shifts in states $\ket{3}$ and $\ket{2}$ with and without atoms in state $\ket{1}$. Solid curves: Predictions (see text).\label{levelscheme} }
\end{figure*}

Recently, we demonstrated new two-field optical techniques~\cite{WuOptControl1, WuOptControl2}, employing EIT~\cite{EITHarris} in the closed channel to control magnetic Feshbach resonances~\cite{ArunOptcontrol, ArunThesis}, Fig.~\ref{levelscheme}a. Our technique~\cite{ArunOptcontrol} tunes the scattering length near a two-photon resonance, where the loss is at a minimum and tunability of scattering length is at a maximum, in contrast to single-field optical methods~\cite{RempeOptControl, ChinMagicOptControl},  where the maximum tunability in the scattering length is associated with maximum loss. Further, our method employs frequency tuning of a few MHz (small compared to the detuning $\approx 1.5$ THz from the atomic resonance), rather than intensity tuning, producing a negligible change in the  net external potential experienced by the atoms. This eliminates the need for a ``magic" wavelength~\cite{ChinMagicOptControl}  and makes our method universally applicable.

Here we report optical tuning of the scattering length in $^6$Li up to $\Delta a\simeq 12\,a_{bg}$, where $a_{bg} = 62\,a_0$, with $a_0$ the bohr radius. Exploiting this wide tunability, we demonstrate spatial control of interactions by creating an interaction ``sandwich", where the central region of the atomic cloud is resonantly interacting $\Delta a> 10\,a_{bg}$ and is surrounded by two weakly interacting regions $\Delta a \simeq 1\, a_{bg}$.

The basic level scheme of our technique is shown in Fig~\ref{levelscheme}a. Optical fields $\nu_1$ (Rabi frequency $\Omega_1$ and detuning $\Delta_1$) and  $\nu_2$ (Rabi frequency $\Omega_1$ and detuning $\Delta_2$), couple the ground molecular states of the singlet potential, $\ket{g_1}$ and $\ket{g_2}$, to the excited state $\ket{e}$, tuning the energy of $\ket{g_1}$ with suppressed optical scattering. The lowest two hyperfine states in $^6$Li, $\ket{1}$ and $\ket{2}$, have an energy-dependent narrow Feshbach resonance (width $\Delta B = 0.1$ G)  at $B_{res} = 543.27$ G~\cite{OHaraNarrowFB}, where the atoms are predominantly in the spin triplet state $\ket{T,k}$, which tunes downward with magnetic field $B$ as $ -2\mu_B\,B$, where $\mu_B\,$ is the Bohr magneton. The triplet continuum $\ket{T, k}$ is coupled to state $\ket{g_1}$ with a second order hyperfine coupling constant $V_{HF}$, which causes the narrow Feshbach resonance. For our experiments, we use $\Omega_1 = 0.5 \, \gamma_e$, $\Omega_2 = 2.2 \, \gamma_e$, where $\gamma_e = 2\pi \times 11.8 $ MHz is the decay rate of the excited molecular state and $\Delta_1 = +\,2\pi \times 19$ MHz. We define the two-photon detuning $\delta = \Delta_2 - \Delta_1$, which is varied by changing the frequency of the $\nu_2$ laser and holding the frequency of $\nu_1$ laser constant. $\delta \equiv 0$ is the two-photon resonance corresponding to minimum loss. For $\delta \equiv 0$, the state $\ket{g_1}$ also returns to its original unshifted position. The state $\ket{g_1}$ is below (above) its unshifted position for $\delta < 0$ ($\delta > 0$).
\begin{figure}[htb]
	\centering
	\includegraphics[width = 3.4 in]{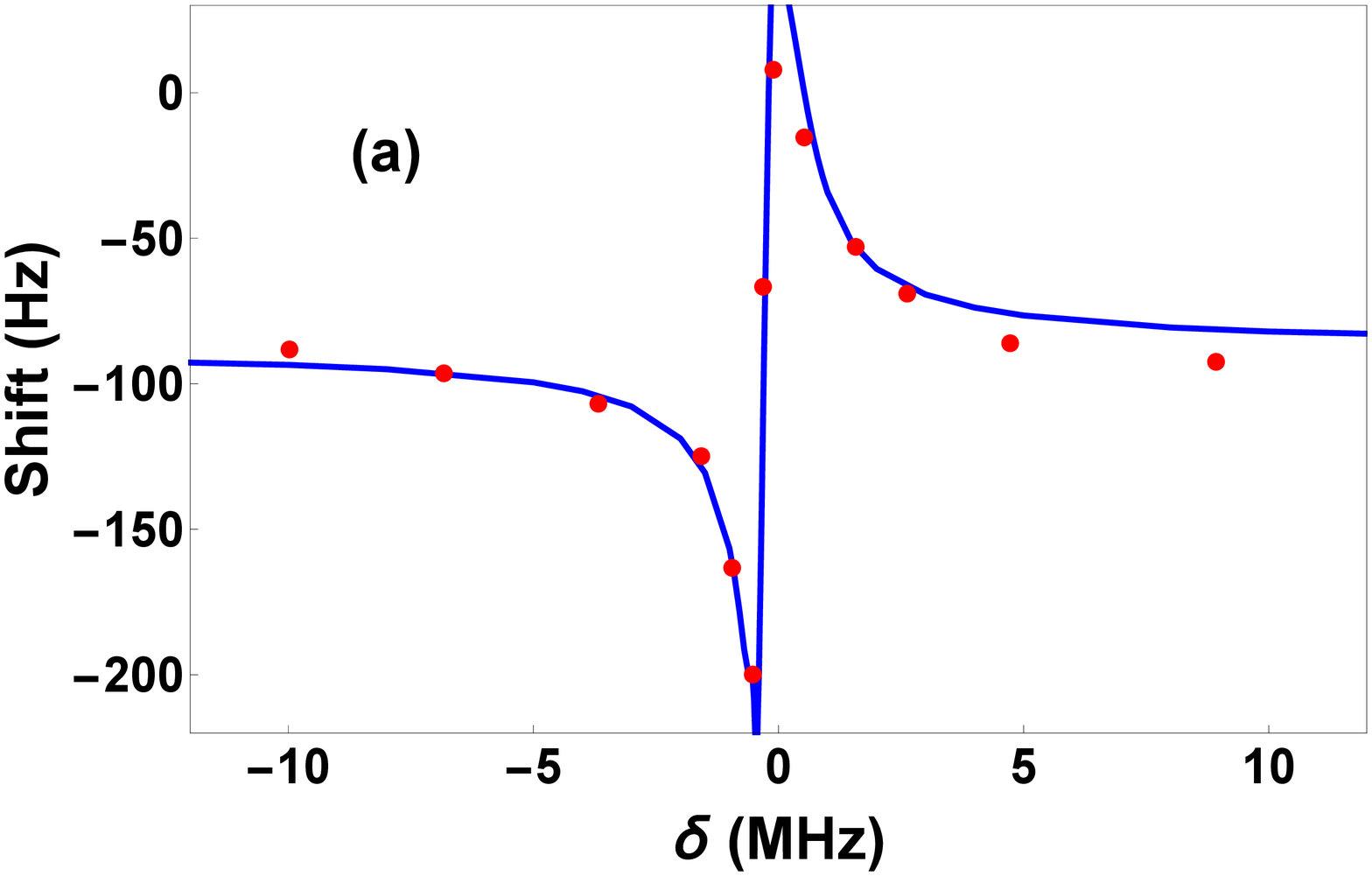}\\
\vspace*{0.125in}
	\includegraphics[width = 3.4 in]{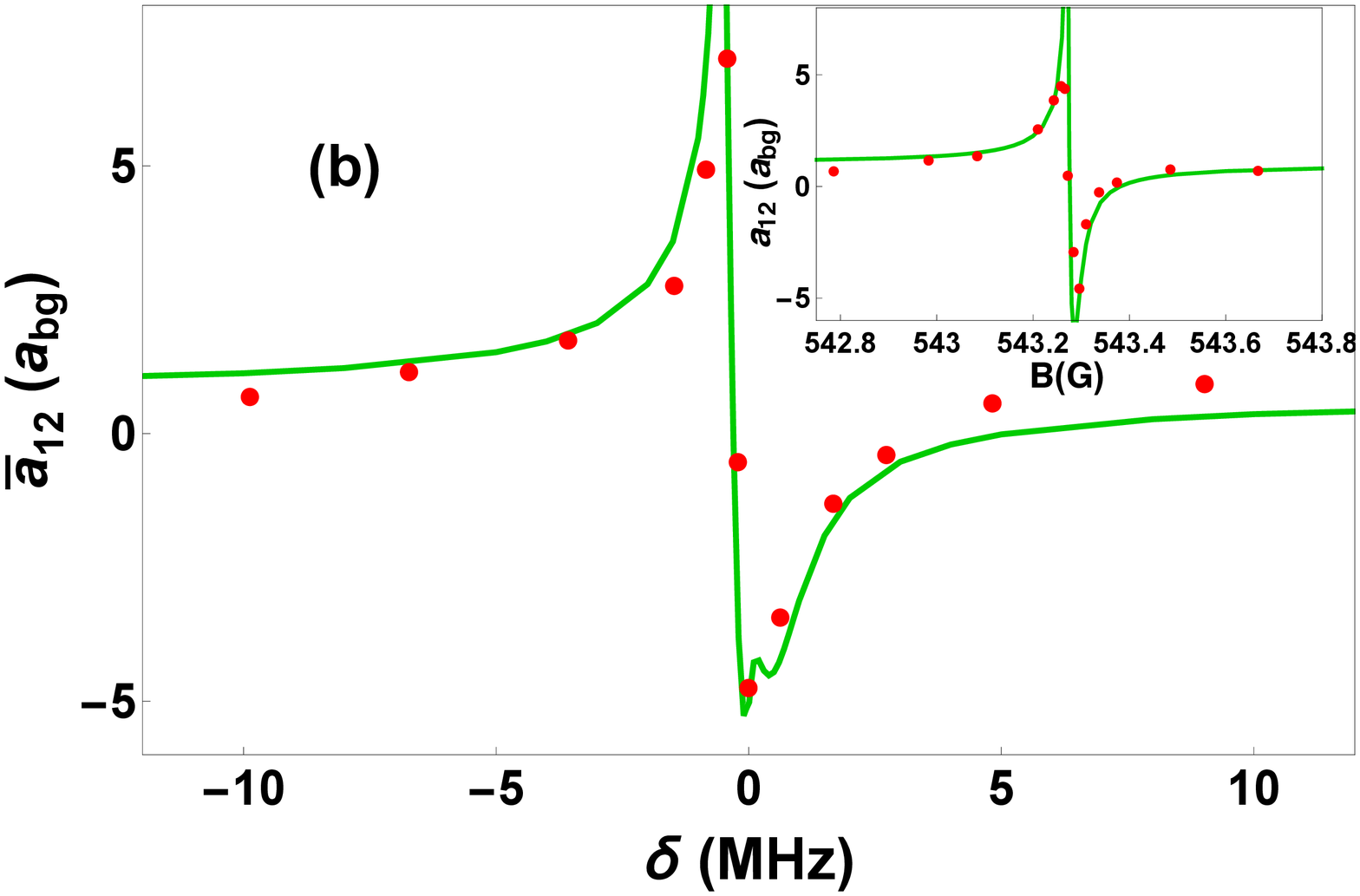}
	\caption{Optical control of the two-body scattering length near the energy dependent narrow Feshbach resonance of $^{6}$Li at 543.28 G (a) Frequency shifts (red dots) in RF spectra as a function of two-photon detuning $\delta$, by changing $\nu_2$ and holding $\nu_1$ constant. $\delta\equiv 0$ denotes the two-photon resonance.  (b) Momentum averaged two-body scattering length $a_{12}$ (red dots) versus $\delta$ determined from the measured frequency shifts. Inset: $a_{12}$ vs magnetic field $B$. Note that optical tuning achieves the same range as magnetic tuning. $a_{bg} = 62\,a_0$. Solid curves: Predictions~\cite{SupportOnline}. \label{opticalscatteringlength}}
\end{figure}

\begin{figure*}[htb]
	\centering
	\includegraphics[width=7 in]{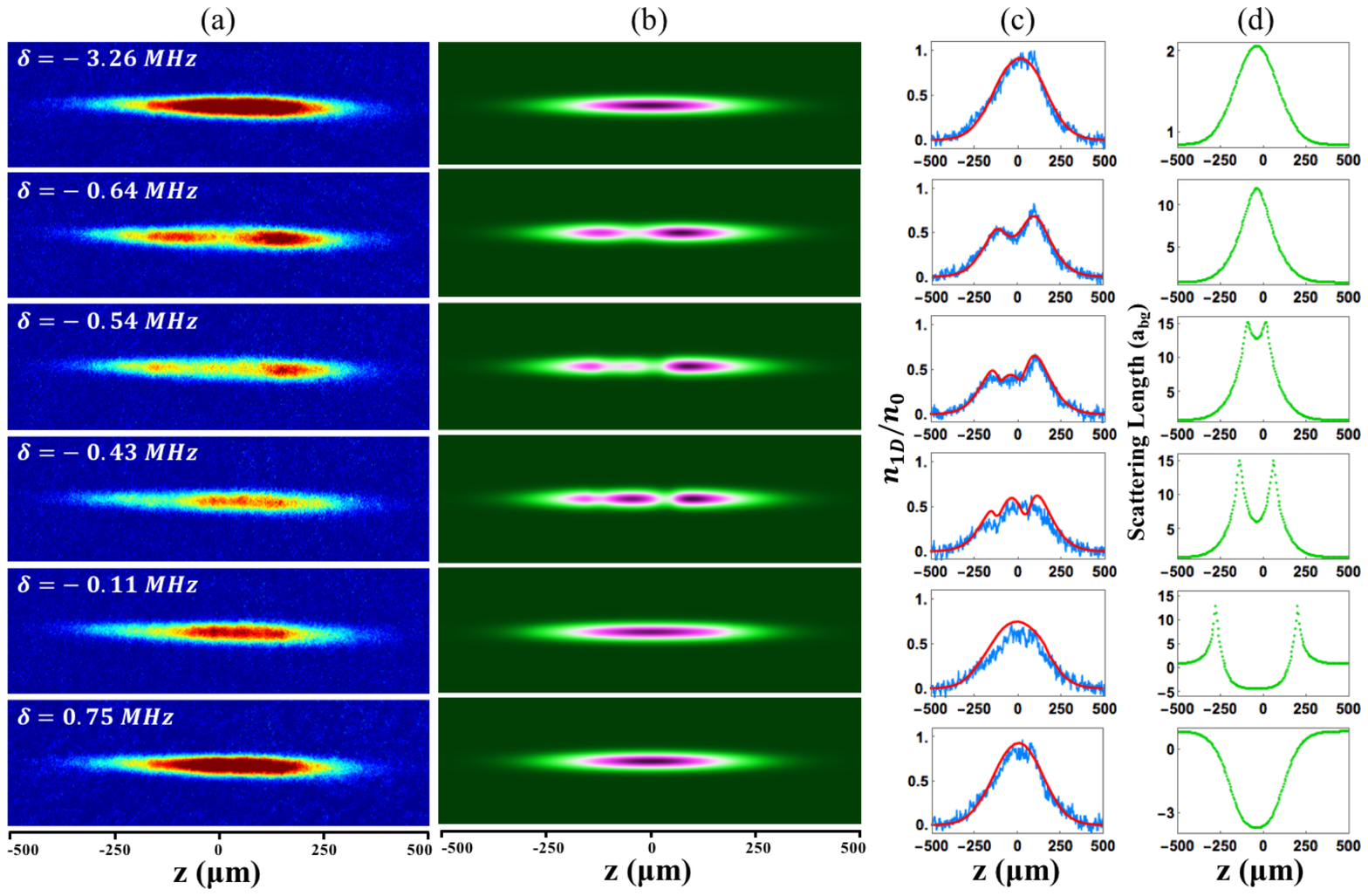}
	\caption{Designer interaction patterns in an ultracold gas of $^{6}$Li atoms versus two-photon detuning $\delta$, with $\delta\equiv 0$ at the two-photon resonance. (a) Measured false color 2D absorption images of atoms transferred from state $\ket{3}$ to state $\ket{2}$ in the presence of atoms in state $\ket{1}$ by applying an RF $\pi$ pulse for 1.2 ms (b) Predicted 2D images using measured parameters~\cite{SupportOnline}; (c) Normalized 1D axial profiles $n_{1D}/n_{0}$, where $n_{0}$ is the peak density with no atoms in state $\ket{1}$. Measured (blue) and calculated (red); (d) Momentum averaged two-body scattering length $a^{opt}_{12}$ used to generate the predicted 2D and 1D spatial profiles.\label{spatial}}
\end{figure*}

We first use radio frequency spectroscopy to demonstrate optical tuning of interactions as a function of the two-photon detuning $\delta$. A trapped cloud of $^6$Li atoms is initially prepared in a mixture of hyperfine states $\ket{1}$ and $\ket{3}$,  Fig.~\ref{levelscheme}b. We ramp the magnetic field to $B = B_{res} + 0.010$ G. We then turn on both optical fields and apply an RF $\pi$ pulse (1.2 ms) that transfers atoms in state $\ket{3}$ to the initially empty state $\ket{2}$~\cite{SupportOnline}. The number of atoms remaining in state $\ket{3}$ is measured by absorption imaging as a function of the radio-frequency. Fig.~\ref{levelscheme}b shows shifted (blue) and unshifted (magenta) RF spectra, obtained with and without atoms in state $\ket{1}$, respectively. The unshifted spectrum  calibrates the magnetic field.  Fig.~\ref{opticalscatteringlength}a (red dots) shows the measured frequency shifts of the RF spectra, as a function  frequency $\nu_2$, holding the frequency $\nu_1$ constant.

The observed frequency shifts are  density dependent and arise from mean-field interactions~\cite{GuptaMeanfield, RegalMeanfield, OHaraNarrowFB} of atoms in states $\ket{3}$ and $\ket{2}$ with atoms in state $\ket{1}$, where  $n_1(\mathbf{r})$ is the density. To understand the data, we calculate the local  transition probability, which depends on the local mean field shift $\Delta\nu$. For two-body scattering, neglecting atom-atom correlations~\cite{MZClockShift,PunkZwergerClockShift},
\begin{equation}
\Delta\nu \text{(Hz)} = \frac{2\,\hbar}{m}\, n_1(\mathbf{r})\, \left[a_{13} - \left\langle a^{opt}_{12}(\nu_2, \Omega_2(z)) \right\rangle\, \right],
\label{freq}
\end{equation}
where $m$ is the atom mass. Here, $\langle a^{opt}_{12} \rangle$ is the  real part of the momentum-averaged, optically controlled, two-body scattering amplitude calculated from the continuum-dressed state model~\cite{ArunOptcontrol,SupportOnline} for the  $\ket{1}-\ket{2}$ narrow Feshbach resonance. Note that $\langle a^{opt}_{12} \rangle$ generally depends on  $\Omega_1$, $\Omega_2$,  $\nu_1$, $\nu_2$, and the magnetic field $B$. The size of the $\nu_2$ beam is comparable to the axial size of the  atom cloud ($z$ - axis)  and hence the Rabi frequency $\Omega_2 (z)$ is z-dependent, enabling spatial control.   In Eq.~\ref{freq}, $a_{13} \approx -267\, a_0$ is the  $\ket{1}-\ket{3}$  two-body scattering length  near 543 G, far from the $\ket{1}-\ket{3}$ broad Feshbach resonance (width $\Delta B \approx 122$ G)  at 690 G~\cite{BartensteinFeshbach, OHaraNarrowFB}. Using the measured Rabi frequencies and RF pulse duration, we compute the  probability for a transition from $\ket{3}$ to $\ket{2}$ and integrate over the phase space distribution of atoms in state $\ket{3}$~\cite{SupportOnline}. The solid blue curve in Fig.~\ref{opticalscatteringlength}a is the predicted frequency shift, which is in excellent agreement with the measurements.

From the measured frequency shifts, we can also determine an average two-body scattering length $\bar{a}_{12}$ for the narrow Feshbach resonance, by assuming $\Delta \nu_{meas}=\frac{2\,\hbar}{m}\,\bar{n}_1\,(a_{13}-\bar{a}_{12})$ and using $\bar{n}_1$ as a fixed fit parameter for all of the data, Fig.~\ref{opticalscatteringlength}b (red dots).
Here, we determine $\bar{n}_1 = 1.5 \times 10^{11}\,{\rm cm}^{-3}$ by fitting the theoretical model for $z=0$, Fig.~\ref{opticalscatteringlength}b green curve, to the measured frequency shifts away from resonance, at $\delta\simeq -10$ MHz. The full solid green curve is the scattering length calculated from the continuum-dressed state model, from which the predicted shifts shown in Fig.~\ref{opticalscatteringlength}a (solid blue curve) is generated.

Fig.~\ref{opticalscatteringlength}b shows that our EIT method tunes the two-body scattering length between $+7\,a_{bg}$ (BEC side of resonance)  and $-5\,a_{bg}$ (BCS side) by changing the frequency $\delta$ by just a few MHz, the same range as obtained by magnetic tuning without optical fields, Fig.~\ref{opticalscatteringlength}b (Inset). In both cases, the tunability is primarily limited by the energy-dependance of the scattering length near the narrow Feshbach resonance, i.e., the large effective range. Furthermore, we note that for the time scale (1.2 ms) used in our optical control experiments, the atom loss due to spontaneous scattering is negligible.

Fig.~\ref{spatial} illustrates spatial control of interactions, using the two-photon detuning $\delta$ as a control parameter. After illuminating the atoms with the $\nu_1$ and $\nu_2$ beams, we apply an RF $\pi$ pulse (1.2 ms) that transfers atoms from state $\ket{3}$ to $\ket{2}$ in the presence of atoms in state $\ket{1}$. The frequency of the RF pulse is chosen to be resonant for $\delta = \pm 10$ MHz, where the entire cloud is weakly interacting. We image the atoms arriving in state $\ket{2}$ as a function of  $\delta$, by varying $\nu_2$ and holding $\nu_1$ constant.

The measured 2D absorption images are shown in Fig.~\ref{spatial}a. The corresponding 1D axial profiles are shown in Fig.~\ref{spatial}c (blue). The transferred fraction of atoms in state $\ket{2}$ depends on the spatially varying, optically controlled $\ket{1}-\ket{2}$ scattering amplitude. Fig.~\ref{spatial}d shows the two-body scattering length $a^{opt}_{12} (z)$ used to generate the predicted 1D spatial profiles (red curves in  Fig.~\ref{spatial}c) and the predicted 2D absorption images in Fig.~\ref{spatial}b.

Excellent quantitative agreement is obtained between the measured (blue) and the calculated (red) 1D axial profiles, Fig.~\ref{spatial}c. The asymmetry in the 1D profiles for $\delta = -0.64$ MHz and $\delta = -0.54$ MHz is due to the off-center position of the $\nu_2$ beam, which is taken into account in generating the calculated 1D profiles.

At $\delta = -0.64$ MHz, we create an interaction ``sandwich," where the central region of the atomic cloud is resonantly interacting with $a_{12} \approx 12\, a_{bg}$ and is enclosed by two weakly interacting regions with $a_{12} \approx 1\, a_{bg}$ (Fig.~\ref{spatial}d). This is  evident from the measured 2D profile in (Fig.~\ref{spatial}a), where the transferred fraction of atoms in the central region of the cloud is heavily suppressed due to the large frequency shift arising from resonant interactions.

 We see that  a small frequency change from $\delta = -0.64$ MHz to $\delta = -0.43$ MHz, inverts the interaction  ``sandwich" by making the central region more weakly interacting than the wings of the atomic cloud, resulting in increased transfer near the center, Fig.~\ref{spatial}a. We also can invert the sign of the interactions between the central and the outer regions of the cloud, Fig.~\ref{spatial}d. For $\delta = -0.11$ MHz, the interactions in the central region become attractive with $a_{12} \approx -5\, a_{bg}$ and the interactions in the wings become repulsive with $a_{12} \approx 10\, a_{bg}$. As $\delta$ is tuned from below the two-photon resonance, $\delta = -3.26$ MHz,  to above the two-photon resonance, $\delta = +0.75$ MHz, the interactions in the central region of the cloud changes sign from repulsive to attractive~\cite{SupportOnline}. We see that a $\delta$ tuning range of just 4 MHz imprints widely different interaction ``designs" on the atomic cloud.

Although spatially varying interactions based on optical techniques have been reported before, previous experiments either suffered from  extremely short (10 $\mu$s) lifetimes~\cite{YamazakiSpatialMod} or limited optical tunability,  $0.15\,a_{bg}$~\cite{ChinMagicOptControl}. Further, all-optical manipulation of spatial interaction profiles has not been previously demonstrated. The two-field EIT method demonstrated here provides a robust, frequency tunable method of spatially manipulating interactions in ultracold atoms and enables temporal control of local interactions, which can be used to study local non-equilibrium thermodynamics.

Our method has broad applications, creating new fields of study in ultracold gases. For example, one can imprint an interaction superlattice, where  interactions between atoms at different lattice sites are independently controlled and manipulated with minimum scattering loss, permitting studies of ``collisionally inhomogenous" systems~\cite{CollisionalInhomeogneous}. Further, a momentum selective extension of our method has been suggested as a means for realizing synthetic Fulde-Ferrell superfluids, where resonant interactions and atom pairing occur at finite momentum, with suppressed optical loss~\cite{OptcontrolCOM, FuldeFerrell}.

\appendix
\newpage
\widetext
\section{Supplemental Material}
\subsection{Experiment}
\label{sec:expt}

We prepare a 50-50 mixture of $^6$Li atoms in the two lowest hyperfine states  $\ket {1}$ and $\ket {2}$. After evaporatively cooling the atoms at 300 G in a CO$_2$ optical trap, we re-raise the trap to 2\% of maximum. The typical temperature is $T = 1.0\, \mu$K and the Fermi temperature is $T_F = 1.4 \, \mu$K. The magnetic field is then ramped to 528 G, where the $\ket {1}-\ket {2}$ mixture is non-interacting. An RF sweep then transfers atoms from state $\ket {2}$ to state $\ket {3}$, resulting in a $\ket {1}-\ket {3}$ mixture. The trap depth is raised to 5\% of the maximum and the magnetic field is ramped to the field of interest, $B  = B_{res} + 0.010$ G, where we measure $B_{res} = 543.27$ G for the narrow Feshbach resonance in $^6$Li. For our optical control experiments, Fig.~1a of the main text, two optical fields $\nu_1$ and $\nu_2$ couple the ground molecular states $\ket{g_1}$ and $\ket{g_2}$ to the excited molecular state $\ket{e}$ of the singlet potential.

To measure frequency shifts in the RF spectra arising from mean field interactions, Fig.~1 of the main text, we initially apply the $\nu_2$ beam with Rabi frequency $\Omega_2$ = 2.1 $\gamma_e$, where $\gamma_e = 2\pi \times$ 11.8 MHz is the decay rate of the excited molecular state. The $\nu_2$ beam creates an non-negligible confinement in the long $z$ direction of the cloud. We wait 50 ms for the atoms to reach equilibrium in the combined potential created by the $\nu_2$ beam and the CO$_2$ laser trap. The $\nu_1$ beam with Rabi frequency $\Omega_1$ = 0.5 $\gamma_e$ and detuning $\nu_1-\nu_{eg_1} = 19$ MHz is then applied. Concurrently, an RF $\pi$ pulse is applied for 1.2 ms, which transfers the atoms from state $\ket {3}$ to state $\ket {2}$. The atoms in state $\ket {3}$ are then imaged after a time of flight of 200 $\mu$s, yielding the frequency shifted RF spectra  (Fig.~1b Blue of the main text). For measuring the unshifted RF spectra (Fig.~1b Magenta of the main text) in the absence of mean-field interactions and to calibrate the magnetic field, we remove the atoms in state $\ket {1}$ by a resonant imaging pulse and then perform RF spectroscopy for the bare $\ket {3}$-$\ket {2}$ transition.

To demonstrate spatial control of interactions, Fig.~3 of the main text,  we repeat the same procedure, and image the atoms arriving in state $\ket{2}$. The 1/e cloud radii are $\sigma_z = 135 \, \mu$m (axial) and $\sigma_r = 7 \, \mu$m (radial). The 1/e intensity radius of the $\nu_1$ beam and $\nu_2$ beam are $w_1 = 530 \, \mu$m and $w_2 = 175 \, \mu$m, respectively. The size of the $\nu_2$ beam is comparable to the axial size $\sigma_z$  of the  atom cloud and hence the Rabi frequency $\Omega_2 (z)$ is z-dependent, enabling spatial control.

We choose the $\ket{g_1}$ and $\ket{g_2}$ states to be the $\ket {v =38}$ and $\ket{v =37}$ ground vibrational state of the singlet potential. We choose $\ket {e}$ to be the $\ket{v'} =64$ excited vibrational state of the singlet potential. The $\nu_1$ beam couples the $\ket {v =38}$ state to the $\ket{v' =64}$ with Rabi frequency $\Omega_1 = 2 \, \pi \times c_1 \sqrt{I_1}$, where $I_1$ is the intensity of the $\nu_1$ beam in $\text{mW}/\text {mm}^2$. We determine $c_1$ by measuring the light induced shifts of the $\ket {v =38}$ state in the absence of the $\nu_2$ beam~\cite{ArunOptcontrol}. The $\nu_2$ beam couples the $\ket{v =37}$ state to the $\ket{v' =64}$ with Rabi frequency $\Omega_2 = 2 \, \pi \times c_2 \sqrt{I_2}$, where $I_2$ is the intensity of the $\nu_2$ beam. We determine $c_2$ by creating a Autler-Townes splitting of the excited state $\ket{e}$ and measuring the absorption spectra for the transition $\ket{T,k}$-$\ket{e}$ near the broad Feshbach resonance~\cite{ArunOptcontrol}. Our measurements of $c_1 = 4.4\,{\rm MHz}/\sqrt{{\rm mW/mm}^2}$ and $c_2 = 1.26\,{\rm MHz}/\sqrt{{\rm mW/mm}^2}$ are in good agreement with the theoretically calculated values $c_1 = 4.77\,{\rm MHz}/\sqrt{{\rm mW/mm}^2}$ and $c_2 = 1.34\,{\rm MHz}/\sqrt{{\rm mW/mm}^2}$, which are based on the molecular potentials~\cite{Cotethesis, CoteJMS1999}.

\subsection{Achieving spatial control of interactions using closed-channel EIT}
The bias B-field is chosen to be $B = B_{res} + 0.010$ G, such that without optical fields, the triplet state $\ket{T, k}$ is below the unshifted energy $\ket{g_1}$, and the interaction is attractive, with $a_{12}<0$. In the presence of optical fields, at the two-photon resonance, $\delta = 0$, the state $\ket{g_1}$ remains at the unshifted position. As the value of $\delta$ is increased from $\delta < 0$ to $\delta \approx 0$, the energy of state $\ket{g_1}$ is optically tuned from below $\ket{T, k}$ to the initial position above $\ket{T, k}$, thereby changing the two-body interaction from repulsive ($a_{12} > 0$) back to attractive ($a_{12} < 0$), Fig.~2b of the main text.

As the energy shift of $\ket{g_1}$ also depends on the spatially varying Rabi frequency $\Omega_2 (z)$, we achieve spatial control of interactions. Increasing $\Omega_2$ creates a downward energy shift of $\ket{g_1}$ for $\delta < 0$, Fig.~3, main text, first five rows, and an upward energy shift of $\ket{g_1}$ for $\delta > 0$, Fig.~3, main text, last row.

\subsection{Theory}
\label{sec:theory}

We begin by finding the probability for an atom, initially in state $\ket{3}$ at position $\mathbf{r}$ with momentum $\mathbf{p}_a$, to make a radio frequency transition to state $\ket{2}$. This atom is immersed in a bath of perturbing atoms in state $\ket{1}$ with 3D density $n_{3D}(\mathbf{r})$ and normalized momentum distribution $W(\mathbf{p}_p)$. Our experiments are performed in the non-degenerate regime, where
\begin{equation}
W(\mathbf{p}')=\frac{1}{\pi^{3/2}p_0^3}\,\exp\left(-\frac{\mathbf{p}'^2}{p_0^2}\right).
\end{equation}
Here, $p_0=\sqrt{2mk_BT}$, $\mathbf{p}'=\mathbf{p}_p$ for the perturbers and  $\mathbf{p}'=\mathbf{p}_a$ for the active atoms.

For a radio frequency pulse of duration $\tau$, and Rabi frequency $\Omega_{RF}$ in Hz, the transition probability is given by
\begin{equation}
P(p_a,\mathbf{r},\nu_{RF})=\frac{\Omega^2_{RF}\sin^2\left[\pi\tau
\sqrt{\Omega_{RF}^2+[\nu_{RF}-\Delta\nu(p_{a},\mathbf{r})]^2}\right]}{\Omega_{RF}^2+[\nu_{RF}-\Delta\nu(p_{a},\mathbf{r})]^2},
\label{eq:prob}
\end{equation}
where the frequency shift in Hz is~\cite{MZClockShift,PunkZwergerClockShift}
\begin{equation}
\Delta\nu(p_{a},\mathbf{r})=\frac{2\hbar}{m}\,n_{3D}(\mathbf{r})\,[a_{13}-a^{opt}(p_{a},z)],
\end{equation}
and
\begin{equation}
 a^{opt}_{12}(p_{a},z)= -\int d^3\mathbf{p}_p\,W(\mathbf{p}_p)\, Re\{f(z,|\mathbf{p}_p-\mathbf{p}_a|/2)\}.
 \label{eq:a12}
\end{equation}
Here, $f$ is the $1-2$ forward scattering amplitude, which depends on the magnitude of the relative momentum $\mathbf{p}=(\mathbf{p}_p-\mathbf{p}_a)/2$ and the position $z$, due to the spatially varying Rabi frequency $\Omega_2(z)$.  Using the relative momentum $\mathbf{p}$ as the integration variable for fixed $\mathbf{p}_a$, we obtain
\begin{equation}
a^{opt}_{12}(p_{a},z)= -\frac{8}{p_a\,p_0\,\sqrt{\pi}}\int_0^\infty dp\,p\,\exp\left(-\frac{4p^2+p_a^2}{p_0^2}\right)\,\sinh\left(\frac{4 p_a p}{p_0^2}\right) Re\{f(z,p)\}.
\end{equation}
We note that generally, the average relative momentum is dependent on the active atom momentum $\mathbf{p}_a$. For s-wave scattering,  we see that $a^{opt}_{12}(p_{a},z)$ depends only on  $p_a=|\mathbf{p}_a|$ after integration over the perturber solid angle $d\Omega_p$.

The one-dimensional z-dependent density of atoms transferred to state $\ket{2}$ is then determined from the $\mathbf{p}_a$-dependent transition probability, by integrating over the momentum distribution of the active atoms, $W(\mathbf{p}_a)$ and over the radial spatial profile of the active atoms, which are initially in state $\ket{3}$,
\begin{equation}
n_2(z,\nu_{RF})=\int_0^\infty 2\pi\rho\, d\rho\, n_{3D}(\rho,z)\,\int d^3\mathbf{p}_a\,W(\mathbf{p}_a)\,P(p_a,\mathbf{r},\nu_{RF}).
\label{eq:1Ddensity}
\end{equation}
Eq.~\ref{eq:1Ddensity} determines the spatial profile of the transferred atoms, Fig.~3 of the main text, which is compared to measurements.

The radio-frequency spectrum is then determined by the total number of atoms transferred to state $\ket{2}$,
\begin{equation}
N_2(\nu_{RF})=\int_{-\infty}^\infty dz\,n_2(z,\nu_{RF}).
\end{equation}
Fig.~1b in the main text shows the normalized fraction of the atoms remaining in state $\ket{3}$,  $N_3(\nu_{RF})/N_0= 1 - N_2(\nu_{RF})/N_0$, where $N_0$ is the initial number of atoms in state $\ket{3}$.

For completeness, we summarize the results of our previous paper~\cite{ArunOptcontrol}, which determines $f(z,p)$, the optically-controlled two-body scattering amplitude, using a continuum-dressed state model,
\begin{equation}
f=\frac{e^{2i\delta_s(k)}-1}{2ik}=\frac{|a_{bg}|}{\tilde{k}\cot\delta_s(\tilde{k})-i\tilde{k}}\,.
\label{eq:scattampl}
\end{equation}
Here, $\hbar k=(\mathbf{p}_p-\mathbf{p}_a)/2$ is the relative momentum and we have defined the corresponding dimensionless relative momentum $\tilde{k}\equiv k|a_{bg}|$ (denoted by $x$ in Ref.~\cite{ArunOptcontrol}), with $a_{bg}$ the background scattering length. The total scattering phase shift $\delta_s(\tilde{k})=\Delta(\tilde{k})+\Phi(\tilde{k})$ is  found from
\begin{eqnarray}
\tilde{k}\cot\delta_s(\tilde{k})&=&\frac{\tilde{k}\cot\Delta(\tilde{k})\,\tilde{k}\cot\Phi(\tilde{k})-
\tilde{k}^2}{\tilde{k}\cot\Phi(\tilde{k})+\tilde{k}\cot\Delta(\tilde{k})}\nonumber\\
&\equiv& q'(\tilde{k})+iq''(\tilde{k}).
\label{eq:q}
\end{eqnarray}
In Eqs.~\ref{eq:scattampl} and~\ref{eq:q},  $\Delta(\tilde{k})$ arises from the background Feshbach resonance, (Eq.~\ref{eq:38.1} below)  while $\Phi(\tilde{k})$  (Eq.~\ref{eq:38.4} below) arises from the optical fields.  For later use, we have defined the real and imaginary parts of $\tilde{k}\cot\delta_s(\tilde{k})$, $q'(\tilde{k})$ and $q''(\tilde{k})$, which determine all of the optically-controlled scattering parameters. Note that $\Phi(\tilde{k})$ and hence $q'(\tilde{k})$ and $q''(\tilde{k})$ are $z$-dependent (suppressed for brevity), due to the spatial variation of the Rabi frequency $\Omega_2(z)$, as noted above. The phase shifts $\Delta$ and $\Phi$ are determined from the scattering state in the continuum dressed state basis~\cite{ArunOptcontrol}.

In the absence of optical fields, the model presented in the Appendix of Ref.~\cite{WuOptControl2} determines the background Feshbach resonance continuum states~\cite{ArunOptcontrol}, yielding
\begin{equation}
\tilde{k}\cot\Delta(\tilde{k})=\frac{\tilde{\Delta}_0-\epsilon\,\tilde{k}^2}{1-{\rm sgn}(a_{bg})(\tilde{\Delta}_0-\epsilon\,\tilde{k}^2)},
\label{eq:38.1}
\end{equation}
where $\tilde{\Delta}_0\equiv(B-B_\infty)/\Delta B$ is the detuning for the magnetic Feshbach resonance and
$\epsilon\equiv E_{bg}/(2\mu_B\Delta B)$, with $E_{bg}=\hbar^2/(ma_{bg}^2)$, $m$ the atom mass, $\mu_B$ the Bohr magneton, and ${\rm sgn}(a_{bg})=\mp\, 1$, for negative or positive $a_{bg}$, respectively.

Including the interaction with both optical fields, we find the asymptotic scattering state in the continuum-dressed basis, which yields the optically-induced phase shift from
\begin{equation}
\tilde{k}\cot\Phi(\tilde{k})=-\frac{\tilde{\delta}_e(\tilde{k})+\frac{\tilde{\Omega}_1^2}{4}\frac{\hbar\gamma_e}{2\mu_B\Delta B}\,S(\tilde{\Delta}_0,\tilde{k})+\frac{i}{2}}{\frac{\tilde{\Omega}_1^2}{4}\frac{\hbar\gamma_e}{2\mu_B\Delta B}\,L(\tilde{\Delta}_0,\tilde{k})},
\label{eq:38.4}
\end{equation}
where
\begin{equation}
L(\tilde{\Delta}_0,\tilde{k})=\frac{1}{(\tilde{\Delta}_0-\epsilon\, \tilde{k}^2)^2+\tilde{k}^2[-{\rm sgn}(a_{bg})+\tilde{\Delta}_0-\epsilon\, \tilde{k}^2]^2}.
\label{eq:38.3}
\end{equation}
The dimensionless detunings and frequencies are given in units of the spontaneous decay rate $\gamma_e$ ($2\pi\times 11.8$ MHz for $^6$Li dimers),
\begin{eqnarray}
&\tilde{\delta}_e(\tilde{k})\equiv\tilde{\Delta}_e(\tilde{k})+\frac{\tilde{\Omega}_2^2}{4\tilde{\delta}(\tilde{k})}\\
&\tilde{\Delta}_e(\tilde{k})\equiv\frac{2\pi\nu_1}{\gamma_e}-\frac{2\mu_B}{\hbar\gamma_e}(B-B_{ref})+\frac{E_{bg}}{\hbar\gamma_e}\tilde{k}^2\nonumber\\
&\tilde{\delta}(\tilde{k})\equiv\frac{2\pi(\nu_2-\nu_1)}{\gamma_e}+\frac{2\mu_B}{\hbar\gamma_e}(B-B_{ref})-\frac{E_{bg}}{\hbar\gamma_e}\tilde{k}^2.\nonumber
\label{eq:39.5}
\end{eqnarray}
Here, we define $\nu_1\equiv0$ to correspond to the field photon $T\rightarrow e$ resonance ($\tilde{\Delta}_e=0$) at the reference magnetic field $B_{ref}$. Similarly, $\nu_2=\nu_1$ is defined to correspond to the two-photon resonance ($\tilde{\delta}=0$) for the $g_1\rightarrow e\rightarrow g_2$ transition. The dimensionless Rabi frequencies are $\tilde{\Omega}_1\equiv\Omega_1/\gamma_e$ for the $g_1\rightarrow e$ transition  and  $\tilde{\Omega}_2\equiv\Omega_2/\gamma_e$ for the $g_2\rightarrow e$ transition.

The second term in the numerator of  Eq.~\ref{eq:38.4} is an $\Omega_1$-dependent frequency shift. For the broad $1-2$ resonance in $^6$Li, where $\epsilon = E_{bg}/(2\mu_B\Delta B)=0.00036<<1$ and $a_{bg}<0$, we obtain
\begin{equation}
S(\tilde{\Delta}_0,\tilde{k})=\frac{\tilde{\Delta}_0+(1+\tilde{\Delta}_0)\tilde{k}^2}
{\tilde{\Delta}_0^2+(1+\tilde{\Delta}_0)^2\tilde{k}^2}\quad\text{for}\,\,\epsilon <<1.
\label{eq:shiftbroad}
\end{equation}
For the  narrow $1-2$ resonance discussed in this paper, where $\epsilon=556>>1$, the shift function is
 \begin{equation}
 S(\tilde{\Delta}_0,\tilde{k})=\frac{1}{\tilde{\Delta}_0-\epsilon\, \tilde{k}^2}\hspace*{0.20in}\text{for}\,\,\epsilon >>1.
 \label{eq:shiftnarrow}
 \end{equation}


\end{document}